\documentclass[aps,pra,twocolumn,showpacs,superscriptaddress]{revtex4}

\usepackage[dvips]{graphicx}
\usepackage{amsmath}
\usepackage{float}

\newcommand{\He}{\rm{H}}

\newcommand{\rmd}{{\rm d}}
\newcommand{\rme}{{\rm e}}

\newcommand{\mbfp}{\mathbf{p}}

\newcommand{\mbfr}{\mathbf{r}}

\newcommand{\mbfv}{\mathbf{v}}

\newcommand{\mbfJ}{\mathbf{J}}

\newcommand{\mbfrho}{\boldsymbol{\rho}}
\newcommand{\mbfsigma}{\boldsymbol{\sigma}}

\newcommand{\Elf}{{\cal E}}
\newcommand{\ELF}{{\boldsymbol{\cal E}}}
\newcommand{\mgf}{{\cal B}}
\newcommand{\MGF}{{\boldsymbol{\cal B}}}

\newcommand{\AVP}{{\boldsymbol{\cal A}}}

\begin{document}

\title{Landau level broadening without disorder, non-integer plateaus without interactions 
-- an alternative model of the quantum Hall effect}
\author{Tobias Kramer}
\affiliation{
Department of Physics,
Harvard University,
17 Oxford Street,
Cambridge, MA 02138, USA.}
\email{tobias.kramer@mytum.de}
\date{January 25, 2006}

\begin{abstract}
I review some aspects of an alternative model of the quantum Hall effect, which is not based on the presence of disorder potentials. Instead, a quantization of the electronic drift current in the presence of crossed electric and magnetic fields is employed to construct a non-linear transport theory. Another important ingredient of the alternative theory is the coupling of the two-dimensional electron gas to the leads and the applied voltages. By working in a picture, where the external voltages fix the chemical potential in the 2D subsystem, the experimentally observed linear relation between the voltage and the location of the quantum Hall plateaus finds an natural explanation. Also, the classical Hall effect emerges as a natural limit of the quantum Hall effect.

For low temperatures (or high currents), a non-integer substructure splits higher Landau levels into sublevels. The appearence of substructure and non-integer plateaus in the resistivity is \textbf{not} linked to electron-electron interactions, but caused by the presence of a (linear) electric field. Some of the resulting fractions correspond exactly to half-integer plateaus.
\end{abstract}

%Quantum Hall effects: Theory and modeling\\

\maketitle

\begin{figure}[b]
\begin{center}
\includegraphics[width=0.99\columnwidth]{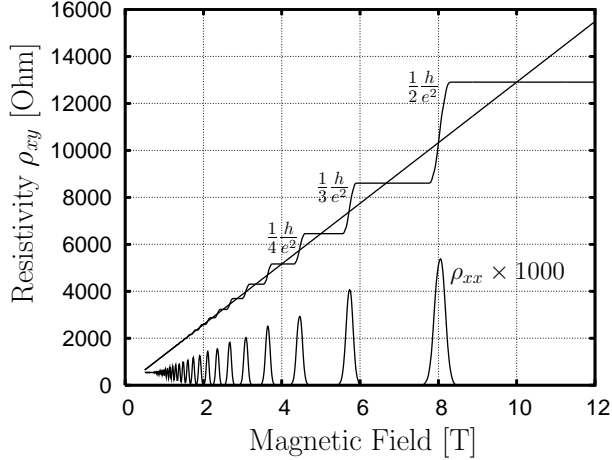}
\end{center}
\caption{
Classical Hall line (straight line) vs.~quantum Hall curve, calculated from \cite{Kramer2005c}.
The QHE leads to a quantized resistance $\rho_{xy}=\frac{1}{i}\frac{h}{e^2}$, $i=1,2,3,\ldots$.
Parameters (references for the values in brackets):
effective mass $m^*=0.1$,
mobility $\mu=17$~m$^{2}$V$^{-1}$s$^{-1}$,
effective $g$-factor $g^*=10$ \cite{EndNoteGFactor},
temperature $T=1$~K, 
current $j_x=1$~Am$^{-1}$,
average carrier density $N_{av}=2.4\times 10^{15}$~m$^{-2}$
(corresponding to a fixed Fermi energy of $E_F=11.6$~meV.)
}\label{fig:hallgraph}
\end{figure}
\begin{figure}[b]
\begin{center}
\includegraphics[width=0.99\columnwidth]{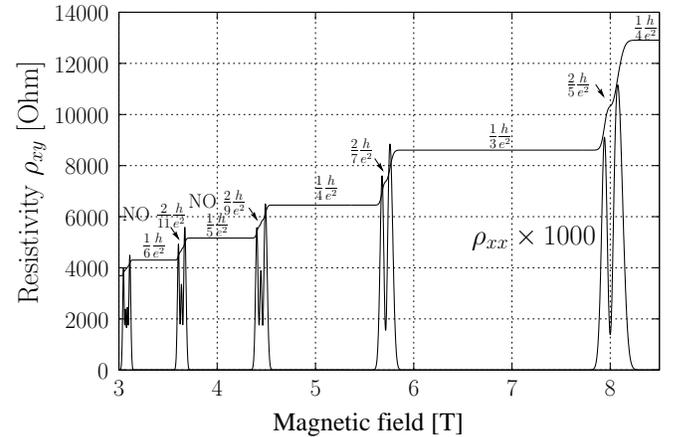}
\end{center}
\caption{
Same as Fig.~\ref{fig:hallgraph}, but at $T=150$~mK. The substructure of the LDOS in higher Landau levels is visible (compare Fig.~\ref{fig:doseb}). Half-filled plateaus exist at $\rho_{xy}=(\frac{2}{5},\frac{2}{7})\frac{h}{e^2}$, however not at $\rho_{xy}=(\frac{2}{9},\frac{2}{11})\frac{h}{e^2}$. Notice that even for $T=0$~K the substructure remains in place, thus giving rise to subdivided Landau levels in a non-interacting particle model.
}\label{fig:hallgraph150}
\end{figure}

\section{The classical Hall effect.}

A purely electric field leads to a uniform acceleration of a charged particle, whereas a purely magnetic field forces the particle on a circular path. The combination of both fields gives rise to the electron drift motion, which is oriented perpendicular to both, electric $\ELF$ and magnetic $\MGF$ fields. Averaging the equation of motions over one cyclotron period $T=2\pi m/(e\mgf)$ yields the drift-velocity:
\begin{equation}\label{eq:DriftVelocity}
\mbfv_d=\frac{1}{T}\int_{t}^{t+T}{\rm d}t'\,
\dot{\mbfr}(t')=(\ELF\times\MGF)/\mgf^2.
\end{equation}
The drift-velocity $\mbfv_d$ is also independent of the initial velocity $\dot{\mbfr}(0)$. 

\begin{figure}[t]
\begin{center}
\includegraphics[width=0.45\columnwidth]{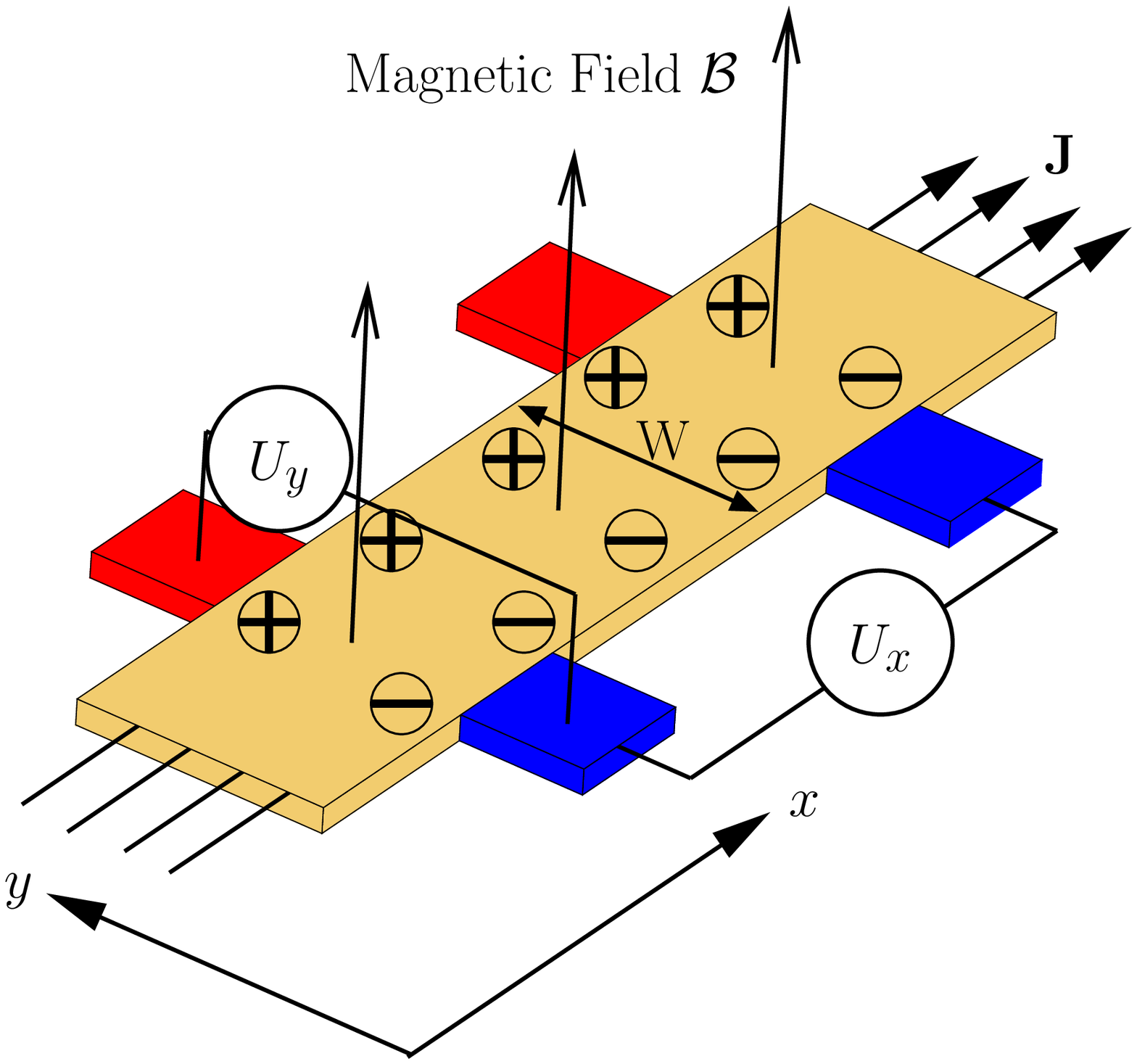}\hfill
\includegraphics[width=0.49\columnwidth]{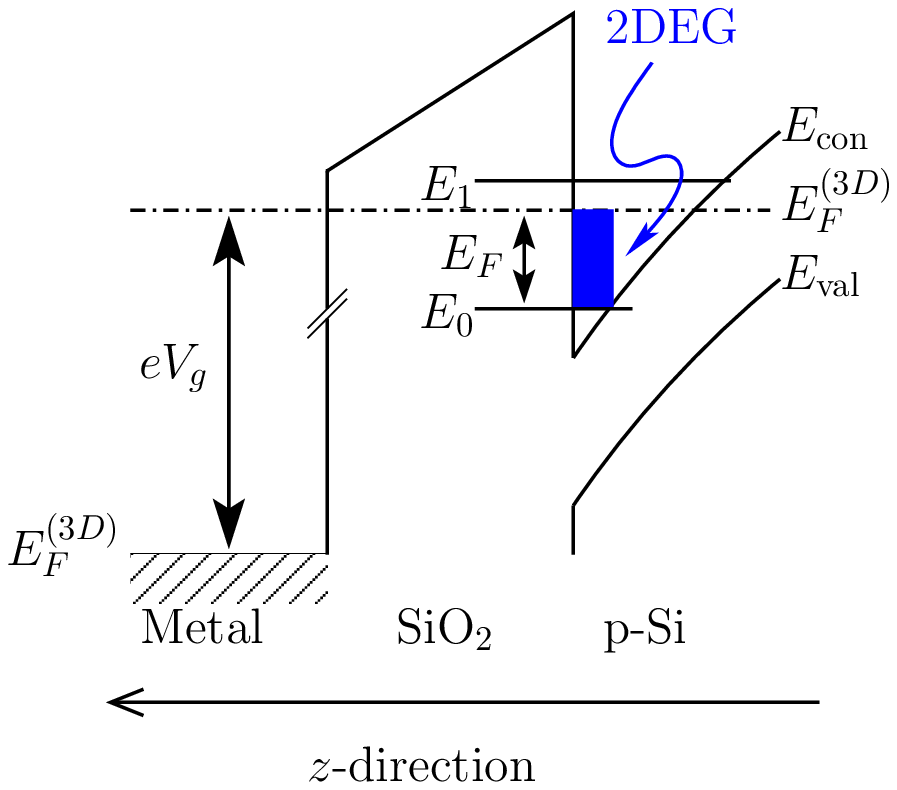}
\end{center}
\caption{
Left panel: Schematic view of a Hall bar. A current $\mathbf{J}$ is flowing through a two-dimensional electron gas (2DEG) in the $x-y$--plane, which is oriented perpendicular to an external magnetic field ${\cal B}$. The deflected electrons at the sample edges produce a Hall voltage $U_y$ over the sample width $W$, which is measured along with the longitudinal voltage drop $U_x$.
Right panel (adapted from \cite{Beenakker1991a}, Fig.~1): Schematic picture of a Metal-Oxide-Semiconductor (MOS) device. The two-dimensional electron gas (2DEG) at the interface between the oxide and the silicon is controlled by applying a gate voltage $V_g$. The gate voltage changes the Fermi energy $E_F^{(3D)}$ of the semiconductor, which in turn couples to the Fermi-energy $E_F$ of the 2DEG. If $E_F<(E_1-E_0)$ holds, the electrons only populate the ground state of the 1D quantum well in $z$-direction that has the eigenenergy $E_0$, which links both Fermi energies via $E_F=E_F^{(3D)}-E_0$.}\label{fig:hallsystem}
\end{figure}

In the following I consider the electronic motion in a two-dimensional subsystem. The orientation of the magnetic field is shown in Fig.~\ref{fig:hallsystem}. The constant drift-velocity has important consequences for the transport of electrons in a solid which is placed in a magnetic field. In a classical Hall experiment deflected electrons form an electric field along the edges of a metal. The conducting electrons propagate in the presence of this electric Hall field, which can be used to determine the carrier-density in the sample \cite{Hall1879a}.
Completely neglecting scattering events, one can extract the basic relation between the classical current $\mbfJ$
\begin{equation}
\mbfJ=N e \mbfv_d,
\end{equation}
($N$ denotes the electron density, $e$ the electronic charge) and the resistivity tensor $\mbfrho$ (or it's inverse, the conductivity tensor $\mbfsigma$) from Ohm's law:
\begin{equation}\label{eq:ConductivityTensor}
\mbfJ=\mbfrho^{-1}\cdot\ELF
\quad\Rightarrow\quad
\mbfrho^{-1}=
\mbfsigma=
\frac{N e}{\mgf}
\left(
\begin{array}{cc}
0 & -1\\
1 & 0
\end{array}
\right).
\end{equation}
The resistivity $\rho_{xy}=\mgf/(N e)$ is proportional to the magnetic field. Notice that the classical Hall effect does in principle not depend on the presence of disorder or scattering processes. The ``electric Hall-field brake'' ensures a constant drift velocity.

\subsection{The quantum Hall effect.}

In contrast to the classical Hall effect, the quantum Hall effect observed by von Klitzing \cite{Klitzing1980a} shows a non-linear variation of the resistivity with the magnetic field.
In the integer quantum Hall effect, the resistivity $\rho_{xy}$ is quantized:
\begin{equation}
\rho_{xy}=\frac{h}{i e^2},\quad i=1,2,3,\ldots
\end{equation}
The conditions for the observation of the quantum Hall effect are low temperatures and very clean samples. 

Interestingly, no standard theory of the integer quantum Hall effect is available. While there exist several models which lead to a quantized resistivity, basic questions remain unanswered: for example, the breakdown of the quantized resistivity above a critical current is documented experimentally, but remains a challenge for most theories.

In the case of the electric field, the difficulty comes from the fact that current theories of the quantum version of the Hall effect are not based on the Hamiltonian of crossed electric and magnetic fields, but rather on the addition of a disorder potential to a purely magnetic field:
\begin{equation}\label{eq:HamiltonDisorder}
H_{\rm lattice,disorder}=
{\left[\mbfp-\frac{e}{c}\AVP(\mbfr)\right]}^2/(2m)
+V_{\rm LD}(\mbfr),
\end{equation}
where $V_{\rm LD}(\mbfr)$ denotes a periodic lattice potential and possibly uncorrelated disorder potentials (which are often assumed to disappear on the average: $\int\rmd\mbfr\, V_{\rm LD}(\mbfr)=0$). 
This Hamiltonian differs from the classical Hall Hamiltonian by the omission of the electric Hall field. The disorder potential becomes an essential part of the description and the appearance of a quantized conductivity is linked to the presence of a fluctuating potential-landscape $V_{\rm LD}$ \cite{Hajdu1994a,KramerB2003a}. Also it cannot sustain an electric field, which would require that the potential landscape is not averaged to zero. Thus for most previous theories of the quantum Hall effect, the electric Hall-field brake is disregarded. In contrast to the classical Hall effect, disorder forms an essential part of the model.

\section{Quantized slopes in the quantum Hall effect.}\label{sec:fluc}
\begin{figure}[t]
\begin{center}
\includegraphics[width=0.5\columnwidth]{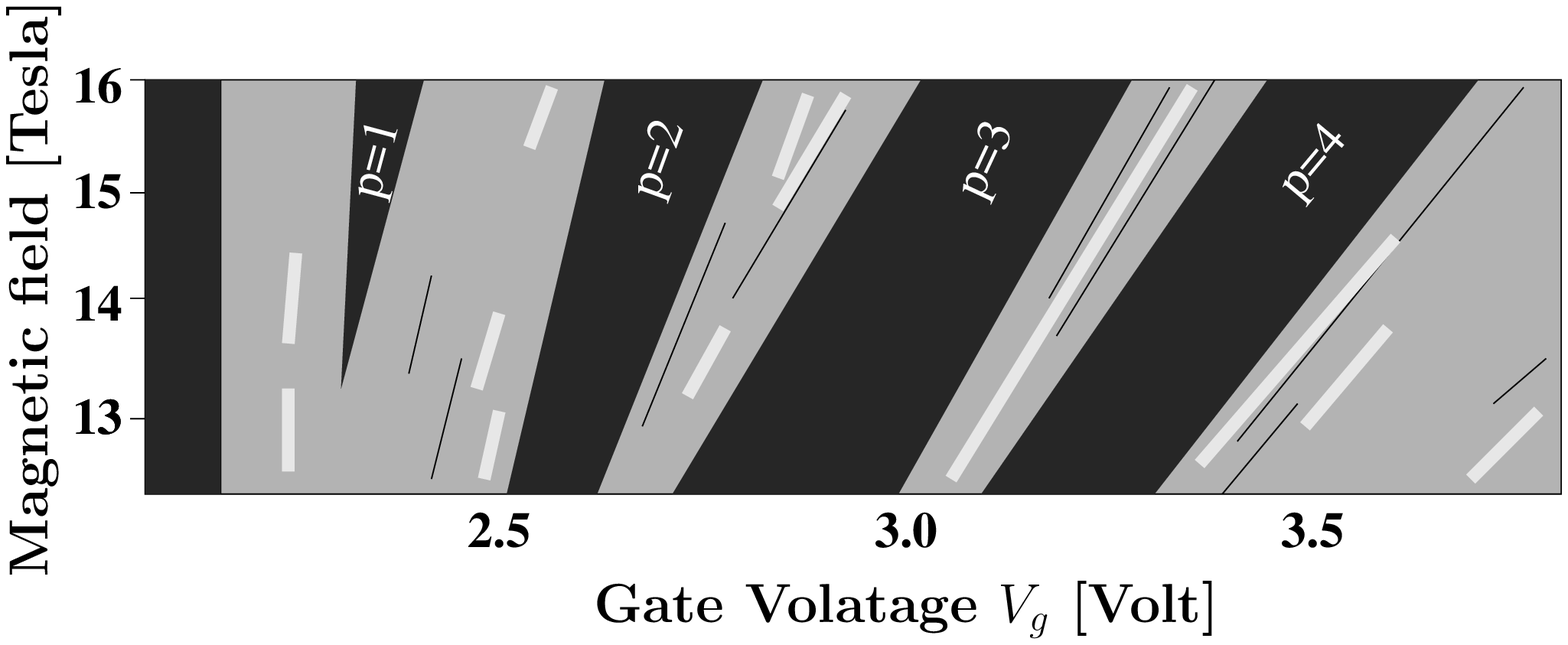}\hfill
\includegraphics[width=0.46\columnwidth]{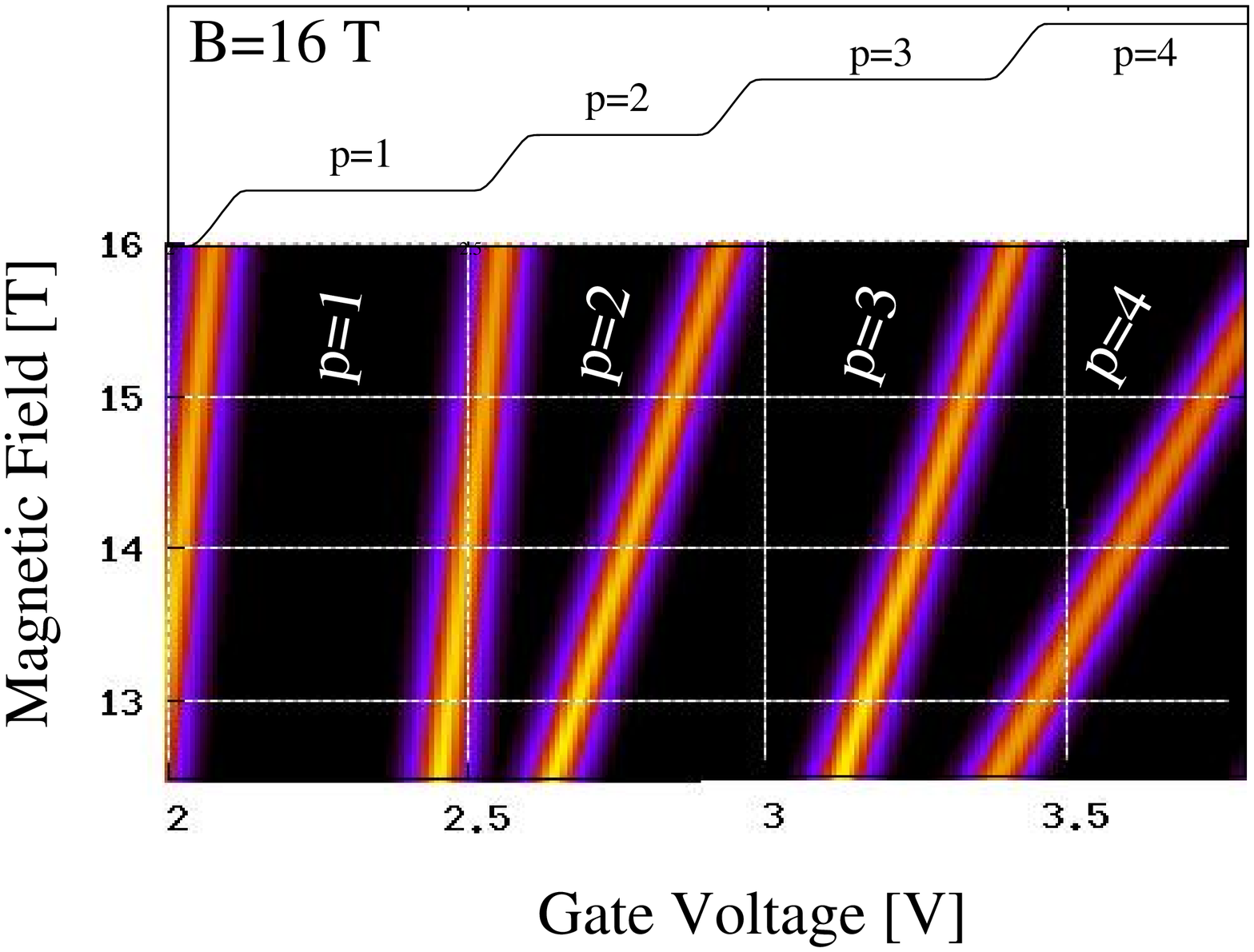}
\end{center}
\caption{Grayscale plot of the conductance $\sigma_{xx}$ as a function of the gate voltage $V_g$ and the magnetic field in a silicon MOSFET. Left panel: A schematic representation of the experimental data obtained by Cobden et al., published in \cite{Cobden1999a}, Fig.~2(a). Right panel: theoretical prediction using \cite{Kramer2005c}, with the following parameters (references for the values in brackets):
transverse effective mass $m^*=0.19$ \cite{Mitin1999a},
mobility $\mu=0.19$~m$^{2}$V$^{-1}$s$^{-1}$ \cite{Cobden1999a},
effective $g$-factor $g^*=5$ \cite{EndNoteGFactor},
valley splitting in silicon $E_{\rm valley}=1.3$~meV \cite{Hirayama2003a},
temperature $T=1.0$~K \cite{Cobden1999a},
$C/e=8.6\times 10^{15}$~m$^{-2}$V$^{-1}$ \cite{Cobden1999a},
$V_{\text{off}}=2.3$~V \cite{Cobden1999a},
$j_y=0.1$~Am$^{-1}$ [assumed]. The location of the plateaus (enumerated by $p$) follows quantized slopes. In the transition region between two $p$'s, the theory shows less structure compared to the experimental result.
}\label{fig:cobden}
\end{figure}
In this section I explore the connection between the density of states (DOS), the Fermi energy, and the number of current carriers. In principle, the number (or density of carriers) is obtained by a convolution integral of two independent quantities: the DOS and the probability of occupation of a quantum state, which is given by the Fermi-Dirac distribution:
\begin{eqnarray}
N(E_F,T)&=&\int_{-\infty}^{\infty}n(E)\,f(E,E_F,T)\,\rmd E\\
f(E,E_F,T)&=&{\left[\rme^{(E-E_F)/(kT)}+1\right]}^{-1}.
\end{eqnarray}
For very low temperatures, the Fermi-Dirac distribution becomes a step-function:
\begin{eqnarray}
N(E_F)&=&\int_{-\infty}^{E_F}n(E)\,\rmd E
\end{eqnarray}
In the absence of external fields, the DOS of a free, non-interacting two-dimensional electron gas (2DEG) is independent of the energy of the state
 \begin{equation}\label{eq:DOSfree}
n_{\text{free}}^{(2D)}(E)=\Theta(E)\frac{m}{2\pi\hbar^2},\quad
\Theta(E)=\left\{
\begin{array}{l}
0\quad E<0\\
1\quad E>0
\end{array}
\right.
\end{equation}
whereas for crossed electric and magnetic fields, the DOS becomes a sum of shifted oscillator densities
\cite{Kramer2003a}, eq.~(20):
\begin{eqnarray}\label{eq:DOSEB}
n_{\ELF\times\MGF}(E) &=& \sum_{k=0}^\infty n_{k,\ELF\times\MGF}(E), \\
n_{k,\ELF\times\MGF}(E) &=&
\frac{{\left[\He_k\left(E_k/\Gamma\right)\right]}^2}{2^{k+1} k! \pi^{3/2}l^2\Gamma} \, \rme^{-E_k^2/\Gamma^2} \, ,
\end{eqnarray}
where $\He_k(x)$ denotes the Hermite polynomial. The level width parameter $\Gamma$ and the energies $E_k$ are given by
\begin{eqnarray}
\Gamma &=& e \Elf_y \sqrt{\hbar/(e\mgf)}\nonumber\\
\omega_L&=&\frac{e\mgf}{2m}\\
\quad E_k&=&E-\Gamma^2/(4\hbar\omega_L)-(2k+1)\hbar\omega_L.\nonumber
\end{eqnarray}

The question how the two-dimensional (quantized) subsystem is coupled to the contacts in an experiments is important for a model of the QHE. In principle, one can think of two possibilities:
\begin{itemize}
\item One can view the subsystem as completely isolated and filled with a fixed number of particles. In this $N={\rm const}$ picture, a change in the underlying DOS (i.e.\ by a change in the magnetic field), yields in principle a change of the energy in the system.
\item On the other hand, a system which is part of an electric circuit can undergo fluctuations in the number of particles, whereas the energy remains fixed.
\end{itemize}
Traditional theories of the QHE try to use an $N={\rm const}$ picture for the current-carrying electrons. However, if one defines the energy of the system to be identical to the last occupied state, problems arise from the absence of available states in the gap between two Landau levels. Another mechanism is needed to ``pin'' the Fermi-energy in between two Landau levels. A commonly used approach is the addition of another kind of density of states, which does not support a current but only provides a non-zero density of states in the gap. This other kind of electrons act as a reservoir and should buffer the otherwise oscillatory Fermi energy.

The alternative model of the QHE \cite{Kramer2005c} follows a different approach: Instead of adding electrons from a reservoir, I propose to treat the QHE system as a system which is part of an electric circuit and is therefore working at a fixed voltage (or Fermi energy) in three dimensions. The two-dimensional subsystem has a fixed voltage difference to the 3D system and therefore has to adjust its number of carriers in order to fulfill the energy conditions of the complete system (see also the discussion in \cite{Lozovoi2001a}).

In this picture, the QHE can be seen as caused by coupling a system with a fixed number of channels to a larger system. The direct coupling of the Fermi-energy of the complete system and the subsystem to external voltages provides a good way to test this picture.

If the Fermi energy is directly determined by a gate voltage $V_g$ (minus a constant offset voltage $V_o$) via
\begin{equation}\label{eq:EfpropVg}
E_F=\alpha (V_g-V_o), 
\end{equation}
it is possible to obtain the intersection points of the (classical) Hall resistivity with the quantized Hall graph. The intersection points are obtained by equating both resistivities for the same Fermi energy $E_F$ 
\begin{eqnarray}
R_{xy}^{cl}=\frac{\mgf}{e N_{\text{av}}} 
&\overset{!}{=}&
R_{xy}^{qm}=\frac{\mgf}{e \int_0^{E_F} n_{\ELF\times\MGF,\uparrow\downarrow}(E,\Elf,\mgf)\,\rmd E},
\nonumber\\\label{eq:rc}
N_{av}&=&\int_0^{E_F}\rmd E\,2\,n_{\text{free}}^{(2D)}(E)=\frac{m^*}{\pi\hbar^2} E_F,
\end{eqnarray}
where $n_{\text{free}}^{(2D)}(E)$ is given by (\ref{eq:DOSfree}) multiplied by two to account for the spin degeneracy and $n_{\ELF\times\MGF}(E,\Elf,\mgf)$ by eq.~(\ref{eq:DOSEB}) with the addition of a spin-splitting (see Sec.~5.5.2 in Ref.~\cite{Kramer2003d}). Note that the intersection point is not necessarily exactly in the middle of a plateau (see Fig.~\ref{fig:hallgraph}).

At the plateaus $R_{xy}^{qm}=\frac{h}{e^2\;i}, \quad i=1,2,3,\ldots$ holds and simultaneously one reaches the intersection point (\ref{eq:rc}) with the classical Hall line $R_{xy}^{qm}=R_{xy}^{cl}$. Therefore the magnetic field values at the intersection points with the quantized resistivity are given by
\begin{equation}\label{eq:Bi}
\frac{h}{e^2 i}=\frac{B}{e N_{av}} \Rightarrow B_i=\frac{h}{e\;i} N_{av}.
\end{equation}
An example for such an intersection point is $B_2=10$~T in Fig.~\ref{fig:hallgraph}. Now it is possible to derive how the magnetic field value of the intersection points changes as a function of the Fermi energy and therefore of the average particle number. Using Eq.~(\ref{eq:Bi}), I obtain for the slopes in GaAs/AlGaAs heterostructures
\begin{equation}
\frac{\partial B_i}{\partial N_{av}}=\frac{h}{e\;i},
\end{equation}
or expressing $\alpha$ in terms of the capacitance $C$ for a Si-MOSFET [$\alpha=C/(n_{\text{free}}^{(2D)}\,e)$]
\begin{equation}
\frac{e}{C}\frac{\partial B_i}{\partial V_g}=\frac{h}{e\;i}.
\end{equation}
These values reflect exactly the experimentally reported quantized slopes (\cite{Cobden1999a}, Eq.~(1), and \cite{Ilani2004a}, p.~329). 
Disorder was deliberately discarded, although it may be important for the observed fine-structure in the experiments. A comparison of the theoretical prediction with experimental results is shown in Fig.~\ref{fig:cobden}. The excellent agreement supports the underlying model of a Fermi energy which is directly proportional to the applied gate voltage, while the actual number of particles may fluctuate about an average value.

Recent experiments trace the evolution of the plateaus as a simultaneous function of the magnetic field $B$ and an applied gate-voltage $V_g$. Experiments have been performed using GaAs heterostructures  \cite{Ilani2004a} (see also \cite{Kramer2005c}) as well as Silicon MOSFET devices \cite{Cobden1999a} (see Fig.~\ref{fig:cobden}). Both experiments confirm the linear law for the plateau location in the $V_g$--$B$-plane.

\section{The role of the electric field in the quantum Hall effect.}\label{sec:electric}
\begin{figure}[t]
\begin{center}
\includegraphics[width=0.49\columnwidth]{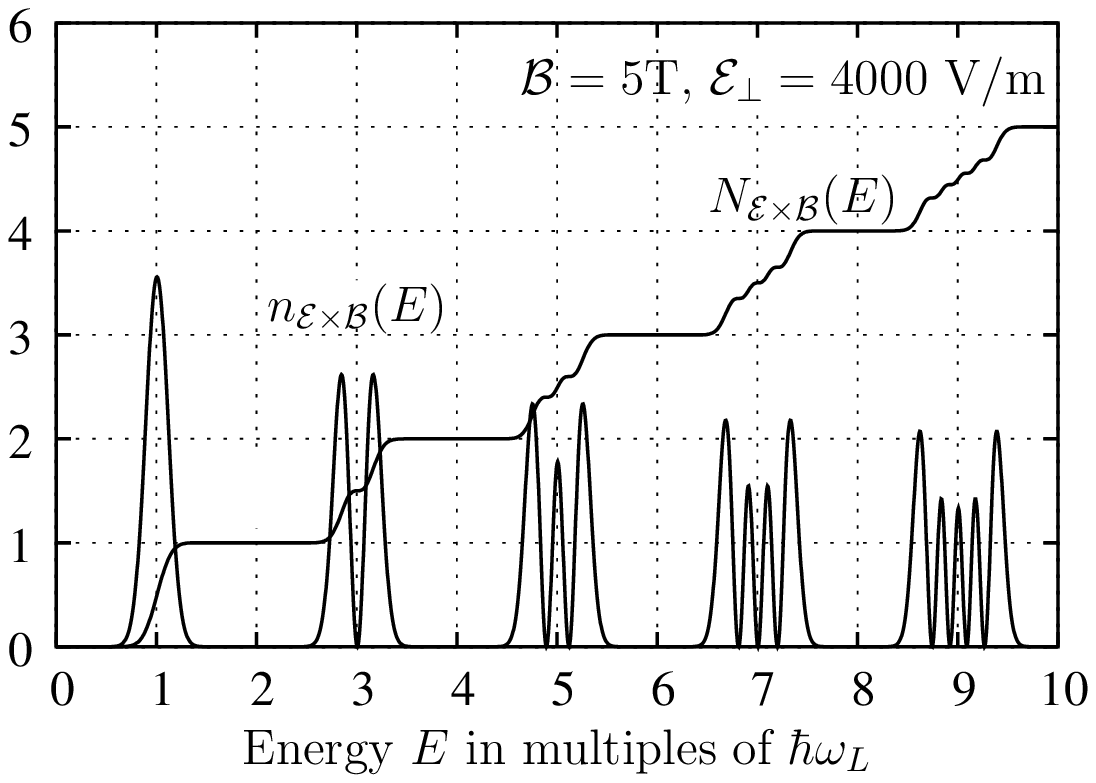}\hfill
\includegraphics[width=0.49\columnwidth]{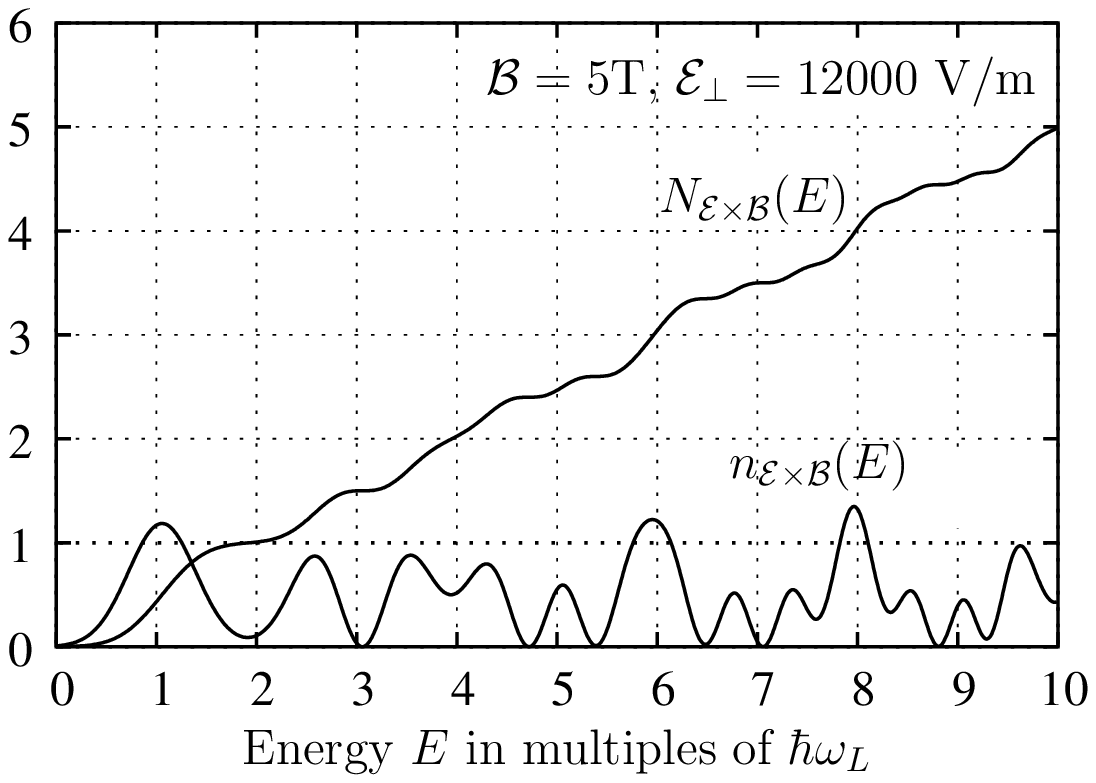}
\end{center}
\caption{Local density of states (LDOS) $n_{{\cal E}\times{\cal B}}(E)$ in crossed electric and magnetic fields, see 
%eq.~(\ref{eq:DOSEB}) and 
\cite{Kramer2003b,Kramer2003a}. $N_{{\cal E}\times{\cal B}}(E)$ denotes the carrier density obtained from $N_{{\cal E}\times{\cal B}}(E)=\int_{-\infty}^{E}n_{{\cal E}\times{\cal B}}(E')\rmd E'$. Note the substructure within Landau levels and the broadening dependent on the electric field value.
% As pointed out in App.~\ref{sec:gdos}, the global density of states does not differ much from the LDOS.
}\label{fig:doseb}
\end{figure}
%%%%
In principle, a microscopic theory of the QHE could work without the presence of the electric field in the basic Hamiltonian, since the electric Hall field is quickly build up as the response of the system to an externally applied voltage. However, to my knowledge, this time-dependent generation of the Hall field is not included in theories of the QHE. Since the steady-state crossed-fields configuration is reached on a short time-scale and the electric field remains present, the field has to be included in the propagation of successive electrons. Interestingly, basic quantities like the local density of states are changed in the presence of an electric field \cite{Kramer2003b,Kramer2003a,Kramer2003d}.

The presence of the electric Hall field does in general not destroy the gaps between two purely magnetic Landau levels, but broadens the Landau levels and imprints a different substructure on each level. These properties are reflected in a non-trivial form of the local density of states (see Fig.~\ref{fig:doseb}) and show the divergence of the quantum Hall effect from a classical electron drift picture: 
\begin{itemize}
\item For emission from a localized contact, the drift depends not only on the field ratio, but also on the kinetic energy of the electrons: for certain energy ranges, localized currents are formed with zero macroscopic flux and the electron propagation is blocked. This is in stark contrast to the classical case, where every electron can participate in the drift motion, independent of its initial (kinetic) energy.
\item Landau-levels are broadened by the electric field in a non-trivial way. Each Landau
level acquires a different substructure and width, dependent on the level number and the electric and magnetic field values (see Fig.~\ref{fig:doseb}).
\item The broadening follows a power law, which leads to a critical Hall field for the breakdown
\begin{equation}\label{eq:crit}
\Elf_{\rm crit}\propto B^{3/2}. 
\end{equation}
\item Higher Landau levels begin to overlap and therefore cannot sustain a quantized transport. Note that there is a natural broadening occuring due to the presence of the Hermite polynomials in eq.~(\ref{eq:DOSEB}).
\end{itemize}
Experiments by Kawaji et al.\ \cite{Kawaji1993a,Kawaji1996a,Shimada1998a}, who studied the QHE and its breakdown as a function of the electric Hall field, are in precise agreement with the theoretical predictions. In fact, the same power law as the theoretically calculated one (see eq.~\ref{eq:crit}) is empirically deduced from the experimental data in \cite{Kawaji1993a}.
\begin{figure}[t]
\begin{center}
\includegraphics[width=0.45\columnwidth]{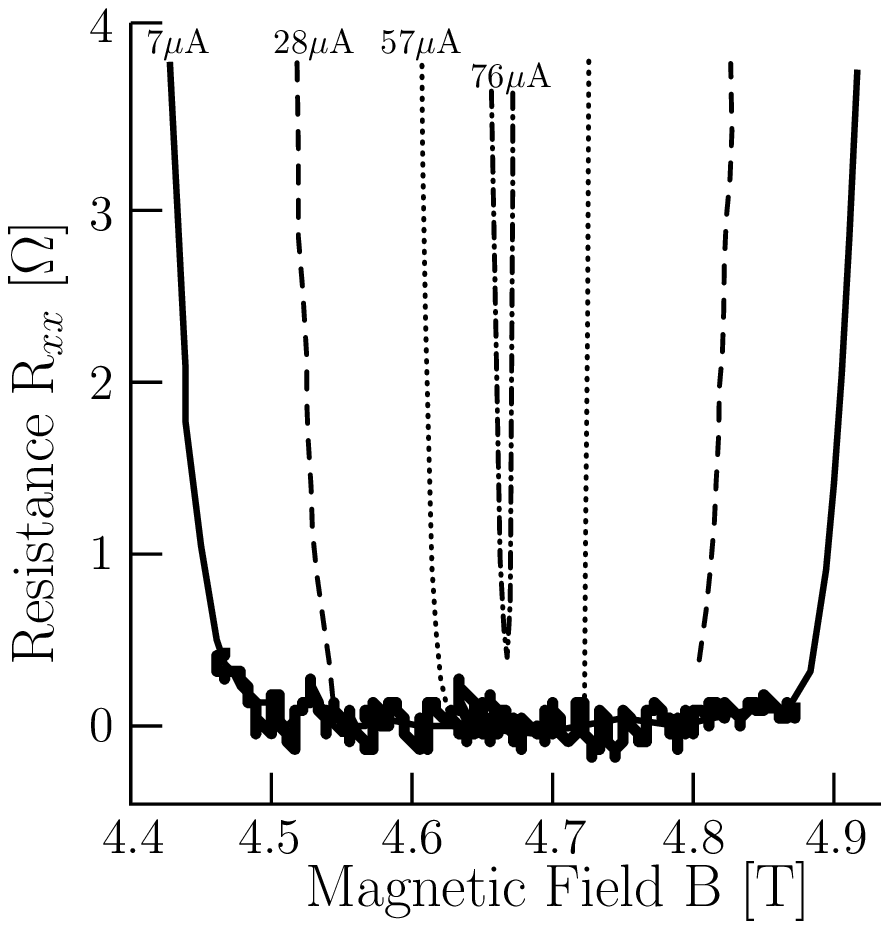}\hfill
\includegraphics[width=0.45\columnwidth]{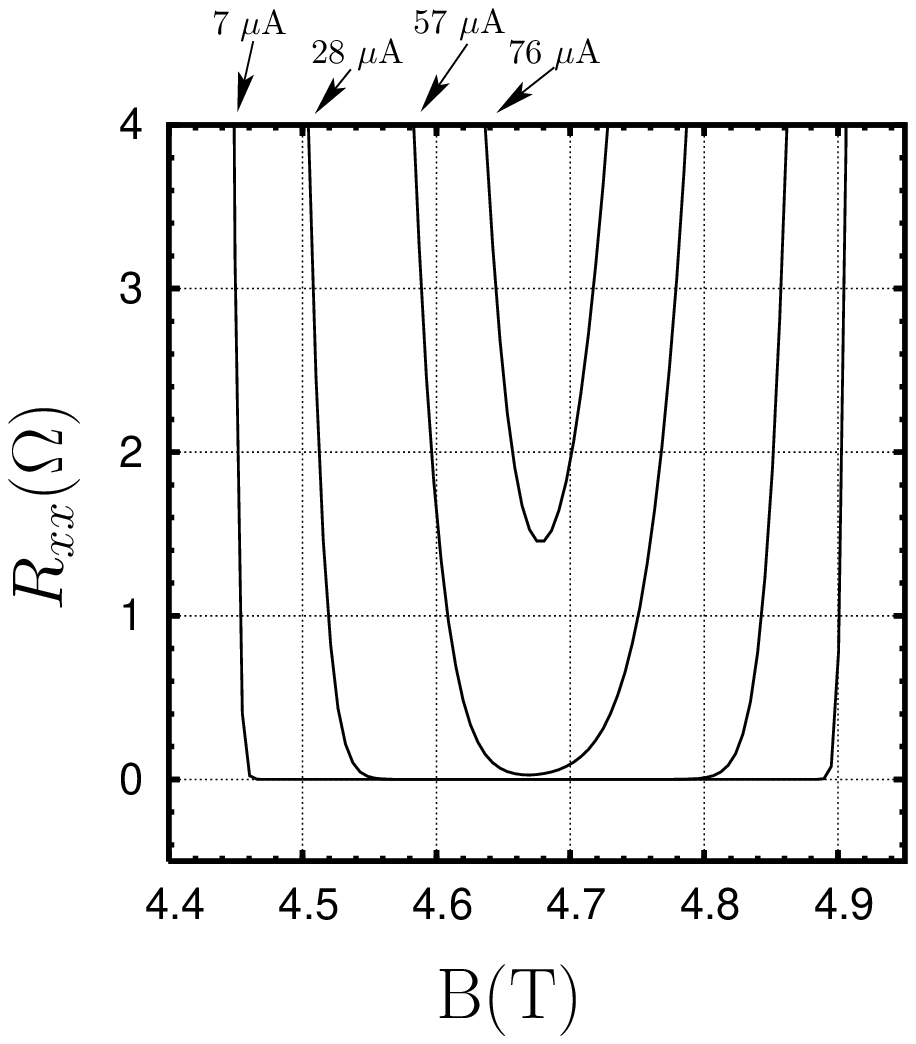}
\end{center}
\caption{Breakdown of the QHE. Diagonal resistance $R_{xx}\propto\sigma_{xx}$ as a function of the magnetic field in the $i=4$ plateau for different currents and therefore electric Hall fields: $j_x=\sigma_{xy}(\mgf,\Elf)\,\Elf_y$. Left panel: A schematic representation of the experimental results obtained by Kawaji, published in \cite{Kawaji1996a}, Fig.~2. Right panel: theoretical prediction using the non-linear expression for the conductivity $\sigma_{xx}(E_F,\Elf_y,\mgf,T,\tau)$ derived in \cite{Kramer2005c}, with the following parameters (references for the values in brackets): Effective mass $m^*=0.1$, scattering time $\tau=1\times 10^{-13}$~s, effective $g$-factor $g^*=12$ \cite{EndNoteGFactor}, temperature $T=1.2$~K, average number of particles $N_{av}=4.5\times 10^{15}$~m$^{-2}$ \cite{Kawaji1996a} (corresponding to a fixed Fermi energy of $E_F=10.7$~meV.) Due to the lack of more experimental data (i.e.\ over a wider magnetic field range), the parameters should be viewed as empirically derived. However, independent of the exact values, the observed power law for the critical Hall field (\ref{eq:crit}) is always reproduced by the theory.}\label{fig:breakdown}
\end{figure}
Also, Kawaji obtains different critical fields for different Landau levels, which is explained by the Landau-level dependent broadening in the theory \cite{Kramer2003a}.  The experimental findings can be explained within the heuristic theory of the Hall conductivity \cite{Kramer2005c}, which goes beyond linear response theories and their assumption of a linear relation between the conductivity and the current. Instead a non-linear relation
\begin{equation}
\mathbf{j}=\boldsymbol{\sigma}(\mgf,\Elf)\cdot\ELF
\end{equation}
is derived. A comparison of the theory and experimental data is shown in Fig.~\ref{fig:breakdown}.

\section{Non-integer plateaus.}

The subdivision of the density of states for higher Landau levels is a suprising result. It is caused by the presence of the electric Hall field. Normally, interactions are invoked to explain a splitting of Landau levels into sublevels. In the present case, no interactions (or disorder) are needed to get a broadening and simultaneously a splitting of Landau levels. Of special interest are half-filled Landau levels, which are stable against variations of the electric field value \cite{Kramer2003a}. In a simple spin-splitting picture, the appearence of spin doubles the appearence of each Landau level due to an energy shift of 
\begin{equation}
\Delta E=\pm \frac{1}{2}g^{*}m\hbar\omega_L.
\end{equation}
If I assume spin-splitted Landau levels (see Fig.~\ref{fig:hallgraph150}), half-integer plateaus are expected at $2/5 \frac{h}{e^2}$ and $2/7 \frac{h}{e^2}$, but not at $2/9 \frac{h}{e^2}$ and $2/11 \frac{h}{e^2}$ (where instead a peak occurs at these values). Interestingly, strong anisotropies have been observed experimentally at these fractions \cite{Lilly1999a}, which warrant a further examination of these fractions, i.e.\ as a function of electric Hall fields.

Notice that the present theory does not predict a subdivision of the lowest Landau level (which is not in line with experiments). A possible explanation is that many-body effects become predominant for low Landau levels at high magnetic fields \cite{Lilly1999a}. Also the fluctuations of the particle number (see Fig.~11 in \cite{Kramer2003d}) are largest at the lowest Landau level, leading to an additional enhancement of interactions at strong magnetic fields.

\section{Conclusions.}
The heuristic theory reviewed in this article has features not contained in conventional theories of the QHE:
\begin{itemize}
\item
It incorporates the electric field in the underlying density of states and yields the classical Hall effect in the limit of strong currents. It explains quantitatively the breakdown of the quantized Hall conductivity. Other theories do not consider the electric Hall field, and are thus unable to explain the (experimentally observed) dependence of the plateau width on the electric Hall field.
\item
The many-body aspect is taken into account by constructing a band model of the QHE, which is filled according to the density of states (DOS) in the presence of the external magnetic field and the electric Hall field. The DOS features gaps in the plateau regions.
\item
The current is calculated in a purely quantum-mechanical way, without using perturbative linear-response theory. The theory shows a sharp contrast between the classical propagation of electrons in crossed electric and magnetic fields emitted from a localized contact and their quantum-mechanical motion \cite{Kramer2003a,Kramer2003d}. 
\item
In contrast to other theories of the QHE, this model allows for fluctuations of the number of carriers about an average value. The coupling between the Fermi energy of the two-dimensional electron gas and the device is provided by a gate voltage (see Fig.~\ref{fig:hallsystem}). The number of carriers is calculated as a function of the gate voltage (and therefore the Fermi energy). Note, that $N(E_F)$ will provide the plateaus, while the average drift velocity is constant. As a result, $N(E_F)$ oscillates as a function of the magnetic field for fixed $E_F$. The gaps in the DOS in perpendicular electric and magnetic fields cause the observed conductivity quantization.
\item Surprisingly, crossed electric and magnetic field induce a substructure in a Landau-level which leads to plateaulike structures at several fractional and nearly fractional values of the conductivity quantum \cite{Kramer2003a}. Although their values match the observed FQHE fractions only partially, it is nevertheless remarkable that a non-interacting particle theory already generates a fractional pattern.
\end{itemize}

\subsection*{Acknowledgments}

I would like to thank the organizers T.~Belyaeva, R.~Bijker, and E.~Mart{\'i}nez~Quiroz for the opportunity to present this work at XXIX Symposium on Nuclear Physics in Cocoyoc, M{\'e}xico. The invitation and hospitality of the Instituto de F{\'i}sica, U.N.A.M., (M.~Moshinsky) and the Instituto de Ciencias Nucleares, U.N.A.M., (A.~Frank) are gratefully acknowledged. I appreciate helpful discussions with P.~Kramer, M.~Kleber, C.~Bracher, and A.~Frank. This work is supported by the Deutsche Forschungsgemeinschaft (grant KR~2889 [Emmy Noether Programme]) and NSEC [E.~Heller, Harvard].

\providecommand{\url}[1]{#1}


\begin{thebibliography}{10}
\bibitem{EndNoteGFactor}
The effective $g$-factor is empirically deduced. I assume a constant $g^{*}$,
  which explains the observed structures very well. Thus a dependency of
  $g^{*}$ on the magnetic field is not needed here.
\end{thebibliography}
\end{document}